\documentstyle[12pt,epsfig]{article}
\newcommand{\ra}{\rightarrow}
\def\be {\begin{equation}}
\def\ee {\end{equation}}

\def\bea {\begin{eqnarray}}
\def\eea {\end{eqnarray}}

\begin{document}

\thispagestyle{empty}
\vskip 15pt

\begin{center}
{\Large {\bf{Measuring isotropization time quark gluon plasma 
from direct photon at RHIC}}}
\renewcommand{\thefootnote}{\alph{footnote}}

\hspace*{\fill}

\hspace*{\fill}

{ \tt{
Lusaka Bhattacharya\footnote{E-mail address:
lusaka.bhattacharya@saha.ac.in}
and
Pradip Roy\footnote{E-mail address:
pradipk.roy@saha.ac.in}
}}\\

\small {\em Saha Institute of Nuclear Physics \\
1/AF Bidhannagar, Kolkata - 700064, INDIA}
\\

\vskip 40pt

{\bf ABSTRACT}
\end{center}

\vskip 0.5cm

We calculate transverse momentum distribution of direct photons from 
various sources by taking into account the initial state momentum 
anisotropy of quark gluon plasma (QGP) and late stage transverse 
flow effects. To evaluate the photon yield from hadronic matter we 
include the contributions from baryon-meson reactions. The total photon 
yield, calculated for various combinations of initial conditions and 
transition temperatures, is then compared with the recent measurement 
of photon transverse momentum distribution by the PHENIX collaboration.
It is shown that due to the initial state anisotropy the photon 
yield from the QGP is larger by a factor of 8-10 than the isotropic 
case in the intermediate $p_T$ regime. It is also demonstrated that the 
presence of such an anisotropy can describe the PHENIX photon data better 
than the isotropic case in the present model. We show that the 
isotropization time thus extracted lies within the range 
$1.5 \geq\tau_{\rm iso} \geq 0.5$ fm/c for the initial conditions used here.

\vskip 30pt

\section{Introduction}

The primary goal of relativistic heavy ion collisions is to create
a new state of matter, called quark gluon plasma  and to study
its properties through various indirect probes. Out of all the
properties of the QGP, the most difficult problem lies in the
determination of isotropization and thermalization time scales
($\tau_{\rm iso}$ and $\tau_{\rm therm}$). Studies on elliptic 
flow (upto about $p_T \sim 1.5 - 2 $ GeV) using ideal hydrodynamics 
indicate that the matter produced in such collisions becomes isotropic 
with $\tau_{\rm iso} \sim 0.6$ fm/c~\cite{Pasi,Hirano,Tannenbaum}. On the 
other hand, using second order transport coefficients with conformal 
symmetry it is found that the isotropization/thermalization time has 
sizable uncertainties~\cite{0805.4552_ref4}. Consequently, there are 
uncertainties in the initial temperature as well. The other uncertain 
parameters are the transition temperature $T_c$, the spatial profile, 
and the effects of flow. Thus it is very necessary to find suitable 
probes which are sensitive to these parameters. Electromagnetic 
probes have been proposed to be one of the most promising tools to 
characterize the initial state of the collisions~\cite{ann,jpr}. Because 
of the very nature of their interactions with the constituents of the 
system they tend to leave the system without much change of their 
energy and momentum. In fact, photons (dilepton as well) can be 
used to determine the initial temperature, or equivalently the 
equilibration time. These are related to the final multiplicity of 
produced hadrons by isentropic expansion of the system formed in heavy 
ion collisions. By comparing the initial temperature with the transition 
temperature from lattice QCD, one can infer whether QGP 
is formed or not. However, it should be remembered that to characterize 
the initial phase through photons one should take into account the 
late stage transverse flow and early stage momentum anisotropy, if any. 
Therefore, photons could be a good probe for early stages of collisions, 
provided the observed flow effects from the late stages of the 
collisions can be understood and modeled properly. The observation 
of pronounced transverse flow in the photon transverse momentum 
distribution has been taken into account in model calculations of 
photon $p_T$ distribution at various beam 
energies~\cite{janejpg,dksepjc,clem,janerap,trenk-hep-ph/0408218}.

Photons are produced at various stages of the evolution process. The 
initial hard scatterings (Compton and annihilation) of partons lead to 
photon production, which we call hard photons. If  
QGP is produced initially, there are  QGP-photons from thermal Compton 
plus annihilation processes. Photons are also produced in different 
hadronic reactions from hadronic matter either formed initially 
(no QGP scenario) or realized as a result of a phase transition 
(assumed to be first order in the present work) from QGP. In addition 
to that there is a large background of photons coming from 
$\pi^0$ and $\eta^0$ decays. The yield of excess photons can be obtained 
if this decay 
contribution is subtracted from the total photon yield.
Photons from hadronic reactions and decays cannot 
be calculated in a model-independent way. The hadronic matter produced 
in heavy ion collisions is usually considered to be a gas of the low 
lying mesons $\pi$, $\rho$, $\omega$, $\eta$ and nucleons. Reactions 
between these as well as the decays of the $\rho$ and $\omega$ were 
considered to be the sources of thermal photons from hadronic 
matter~\cite{ann,kap,song}. We also add the contributions from reactions 
involving baryons as these are found~\cite{rap2} to be comparable to 
that from the meson-meson reactions.

It is to be noted that while estimating photons from QGP, it is assumed that 
the matter formed in the relativistic heavy ion collisions 
is in thermal equilibrium. The measurement of elliptic flow parameter 
and its theoretical explanation also support this assumption. On the 
contrary, perturbative estimation suggests the slower thermalization 
of QGP~\cite{PRC75_ref2}. However, recent hydrodynamical 
studies~\cite{0805.4552_ref4} have shown that due to the poor 
knowledge of the initial conditions there is a sizable amount of 
uncertainty in the estimate of thermalization or isotropization time. 
It is suggested that (momentum) anisotropy driven plasma instabilities 
may speed up the process of isotropization~\cite{plb1181993} and in that 
case one is allowed to use hydrodynamics for the evolution of the matter. 
However, instability-driven isotropization is not yet proved at RHIC 
and LHC energies.

Earlier works~\cite{janejpg,dksepjc,turbide_gale_fro_heinz} on 
photon production assume isotropy from the very beginning, i. e. 
$\tau_{\rm iso} = \tau_i $ (QGP formation time). In view of the 
absence of a theoretical proof behind the rapid thermalization and the 
uncertainties in the hydrodynamical fits of experimental data, such an 
assumption may not be justified. Hence in stead of equating the 
thermalization/isotropization time to the QGP formation time, in this 
work, we will introduce an intermediate time scale 
(isotropization time, $\tau_{\rm iso}$) to study the effects of early 
time momentum-space anisotropy on the total photon yield and compare 
it with the PHENIX photon data~\cite{adler1,adler2,phenix08}.
In the present model the space-time evolution, during the interval 
$\tau_i < \tau <\tau_{\rm iso}$, is  modeled 
as in Ref.~\cite{mauricio}. For the evolution from $\tau_{\rm iso}$ 
to $\tau_{F}$ (freeze-out time) we use (1+2)$d$ ideal hydrodynamics. 

Recently, it has been shown in 
Ref.~\cite{mauricio} that for fixed initial conditions, the introduction 
of a pre-equilibrium momentum-space anisotropy enhances high energy 
dileptons by an order of magnitude.  In case of photon transverse momentum
distribution similar results have been reported for
various evolution scenarios~\cite{lusakaprc}.
The model in Ref.~\cite{mauricio} assumes two time 
scales: the QGP formation time, $\tau_i$, and the isotropization time,
$\tau_{\rm iso}$, which is the time when the system becomes isotropic 
in momentum space. Immediately after the formation of QGP at $T_i$
and $\tau_i$, the system
can be assumed to be isotropic~\cite{5of4552}. Subsequent rapid
expansion of the matter along the beam direction causes faster cooling in
the longitudinal direction than in the transverse
direction~\cite{PRC75_ref2}. As a result, the system becomes anisotropic
with $\langle{p_L}^2\rangle << \langle{p_T}^2\rangle$ in the local rest frame. 
At some later time when the effect of parton interaction rate 
overcomes the plasma expansion rate, the system returns to the 
isotropic state again (at $\tau_{\rm iso}$) and remains isotropic for the 
rest of the period.

The plan of the paper is the following. In the next section
we will discuss the mechanisms of
photon production from various possible sources and
the space-time evolution of the matter very briefly.
Section 3 is devoted  to describe the results for
various initial conditions and we summarize in section 4.

\section{Formalism}
\subsection{Photon rate : Anisotropic QGP}

The lowest order processes for photon emission from QGP are the
Compton ($q ({\bar q})\,g\,\rightarrow\,q ({\bar q})\,
\gamma$) and the annihilation ($q\,{\bar q}\,\rightarrow\,g\,\gamma$)
processes. 
The rate of photon production from 
anisotropic plasma due to Compton and annihilation processes has been 
calculated in Ref.~\cite{prd762007}. The soft contribution is calculated by
evaluating the photon polarization tensor for an oblate momentum-space
anisotropy of the system where the cut-off scale is fixed at 
$k_c \sim  \sqrt g p_{hard}$. Here $p_{hard}$ is a hard-momentum scale 
that appears in the distribution functions. 

The differential photon production rate for $1+2\to3+\gamma$ processes in 
an anisotropic medium is given by~\cite{prd762007}: 
\begin{eqnarray} 
E\frac{dN}{d^4xd^3p}&=& 
\frac{{\mathcal{N}}}{2(2\pi)^3} 
\int \frac{d^3p_1}{2E_1(2\pi)^3}\frac{d^3p_2}{2E_2(2\pi)^3}
\frac{d^3p_3}{2E_3(2\pi)^3}
f_1({\bf{p_1}},p_{\rm hard},\xi)f_2({\bf{p_2}},p_{\rm hard},\xi) \nonumber\\
&\times&(2\pi)^4\delta(p_1+p_2-p_3-p)|{\mathcal{M}}|^2 
[1\pm f_3({\bf{p_3}},p_{\rm hard},\xi)]
\label{photonrate}
\end{eqnarray}
where, $|{\mathcal{M}}|^2$ represents the spin averaged matrix element
squared for one of those processes which contributes to the photon rate
and ${{\mathcal N}}$ is the degeneracy factor of the corresponding
process. $\xi$ is a parameter controlling the strength of the anisotropy 
with $\xi > -1$. $f_1$, $f_2$ and $f_3$ are the anisotropic 
distribution functions of the medium partons and will be 
discussed in the following. Here it is assumed that the infrared 
singularities can be shielded by the thermal masses for the 
participating partons. This is a good approximation at short times  
compared to the time scale when plasma instabilities start to play 
an important role. 

The anisotropic distribution function can be obtained~\cite{stricland} 
by squeezing or stretching an arbitrary isotropic distribution function 
along the preferred direction in momentum space,
\begin{eqnarray}
f_{i}({\bf k},\xi, p_{hard})=f_{i}^{iso}(\sqrt{{\bf k}^{2}+\xi 
({\bf k.n})^{2}},p_{hard})
\label{dist_an}
\end{eqnarray}
where ${\bf n}$ is the direction of anisotropy. It is important to
notice that $\xi > 0$ corresponds to a contraction of the 
distribution function in the direction of 
anisotropy and $-1 < \xi < 0 $ corresponds to a stretching in the
direction of anisotropy. In the context of relativistic
heavy ion collisions, one can identify the direction of anisotropy with 
the beam axis along which the system expands initially. The hard momentum
scale $p_{hard}$ is directly related to the average momentum of the 
partons. In the case of an isotropic QGP, $p_{hard}$ can be identified 
with the plasma temperature ($T$). 

\subsection{Photon rate : Isotropic case} 

As mentioned earlier the QGP evolves  hydrodynamically from 
$\tau_{\rm iso}$ onwards. In such case the distribution functions become 
Fermi-Dirac or Bose-Einstein distributions. The photon emission rate, 
in isotropic case, from Compton 
($q ({\bar q})\,g\,\rightarrow\,q ({\bar q})\,\gamma$) and annihilation 
($q\,{\bar q}\,\rightarrow\,g\,\gamma$) processes has been 
calculated from the imaginary part of the photon self-energy by 
Kapusta et al.~\cite{kap} in the 1-loop approximation. However, it has 
been shown by Auranche et al.~\cite{aur1} that the two loop contribution 
is of the same order as the one loop due to the shielding of infra-red 
singularities. The complete calculation upto two loop was done by 
Arnold et al.~\cite{arnold} and the rate is given by
\begin{equation}
\frac{dN}{d^4xd^3p} = \frac{1}{(2\pi)^3}\,{\cal A}(E,T)\,
\left(\ln[T/m_q(T)]+\frac{1}{2}\ln(2E/T)+C_{\mathrm{tot}}
(E/T)\right),
\end{equation}
where $E=p$ and $m_q^2(T)=4\pi\alpha_s T^2/3$ and ${\cal A}$ is the
leading log coefficient given by
\begin{equation}
{\cal A}(E,T) = 2\,\alpha\,N_c\,\sum_i\,q_i^2\,\frac{m_q^2(T)}{E}\,
f_D(E)
\end{equation}
and
\begin{equation}
C_{\mathrm{tot}} = C_{2\ra 2}(E/T)+C_{\mathrm{brems}}(E/T)
+C_{\mathrm{aws}}(E/T)
\end{equation}
containing the dependence of the specific photon production processes.
These are parameterized as follows:
\bea
C_{2\ra 2}&=&0.04(E/T)^{-1}-0.3615+1.01 \exp(-1.35E/T)
\nonumber\\
C_{\mathrm{brems}}+C_{\mathrm{aws}}&=&
\sqrt{1+\frac{1}{6}N_f}\,
\left(\frac{0.548\ln[12.28+T/E]}{(E/T)^{3/2}}
\right.\nonumber\\
&&\left.+\frac{0.133E/T}{\sqrt{1+(E/T)/16.27}}\right)
\eea
%

\subsection{Photon production rate from hot hadronic matter}

First we shall consider photon emission from reactions of the type 
$M\,M\,\ra\,M\,\gamma$, where $M$ generically denotes the low lying 
mesons. As mentioned earlier these type of reactions are thought to be 
the only sources of photons from hadronic matter and already a 
substantial amount of work has been done
~\cite{ann,kap,song,npa98,npa99,turbide} along this line. 

We follow the calculations done in Ref.~\cite{turbide}
where  convenient parameterizations have been given for the reactions
considered. These parameterizations will be used while doing the space-time
evolution to calculate the photon yield from meson-meson reactions.
The photon emission rate (static) from reactions of the type 
$B\,M\,\ra\, B\,\gamma$ ($B$ denotes baryon) has been 
calculated in Ref.~\cite{rap2}. It is 
shown that this contribution is not negligible compared to that
meson-meson reactions. To evaluate 
photon rate due to nucleon (and antinucleon) scattering from 
$\pi$, $\rho$, $\omega$, $\eta$ and $a_1$ mesons in the thermal bath we 
use the phenomenological interactions described in Ref.~\cite{rap2}.

\subsection{Hard Photons}

Besides the thermal photons from QGP and hadronic matter we also
calculate photons from initial hard scattering from the reaction of the 
type $h_A\,h_B\,\rightarrow\,\gamma\,X$ using perturbative QCD. We include
the transverse momentum broadening in the initial state 
partons~\cite{wong,owens}.
The cross-section for this process can then be written in terms of
elementary parton-parton cross-section multiplied by the partonic flux
which depends on the parton distribution functions (PDF) for which we take CTEQ
parameterization~\cite{cteq}.
A phenomenological factor $K$ is used to take into account the higher
order effects.
\begin{figure}
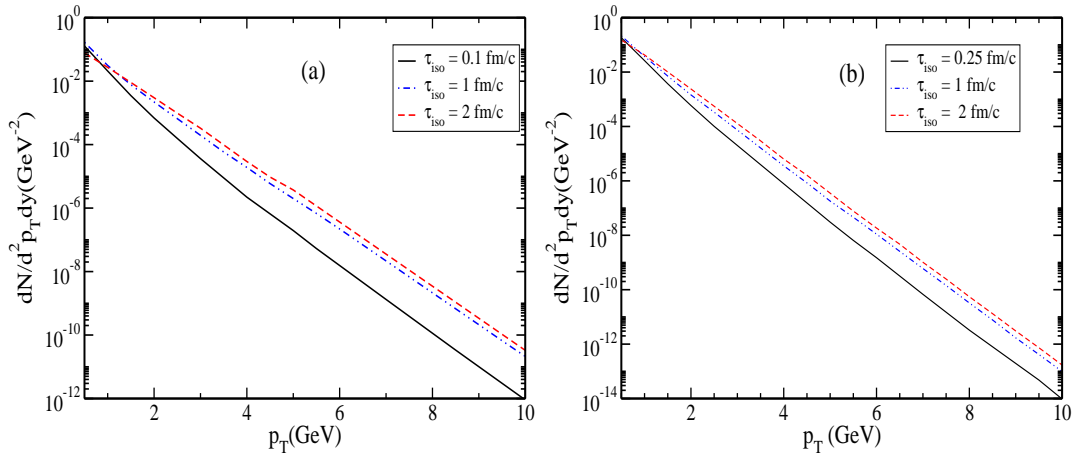

\begin{center}
\epsfig{file=YT440.eps,width=7cm,height=6cm,angle=0}
\epsfig{file=YT350.eps,width=7cm,height=6cm,angle=0}
\end{center}
\caption{(Color online) Medium photon spectrum, $dN/d^2p_Tdy$, at $y=0$ 
for the {\em free-streaming interpolating} model ($\delta = 2$) 
for three different values of isotropization time, $\tau_{iso}$, with 
initial conditions, (a) Set-I and (b) Set-III.}
\label{fig1}
\end{figure}

\subsection{Space time evolution}
\begin{figure}
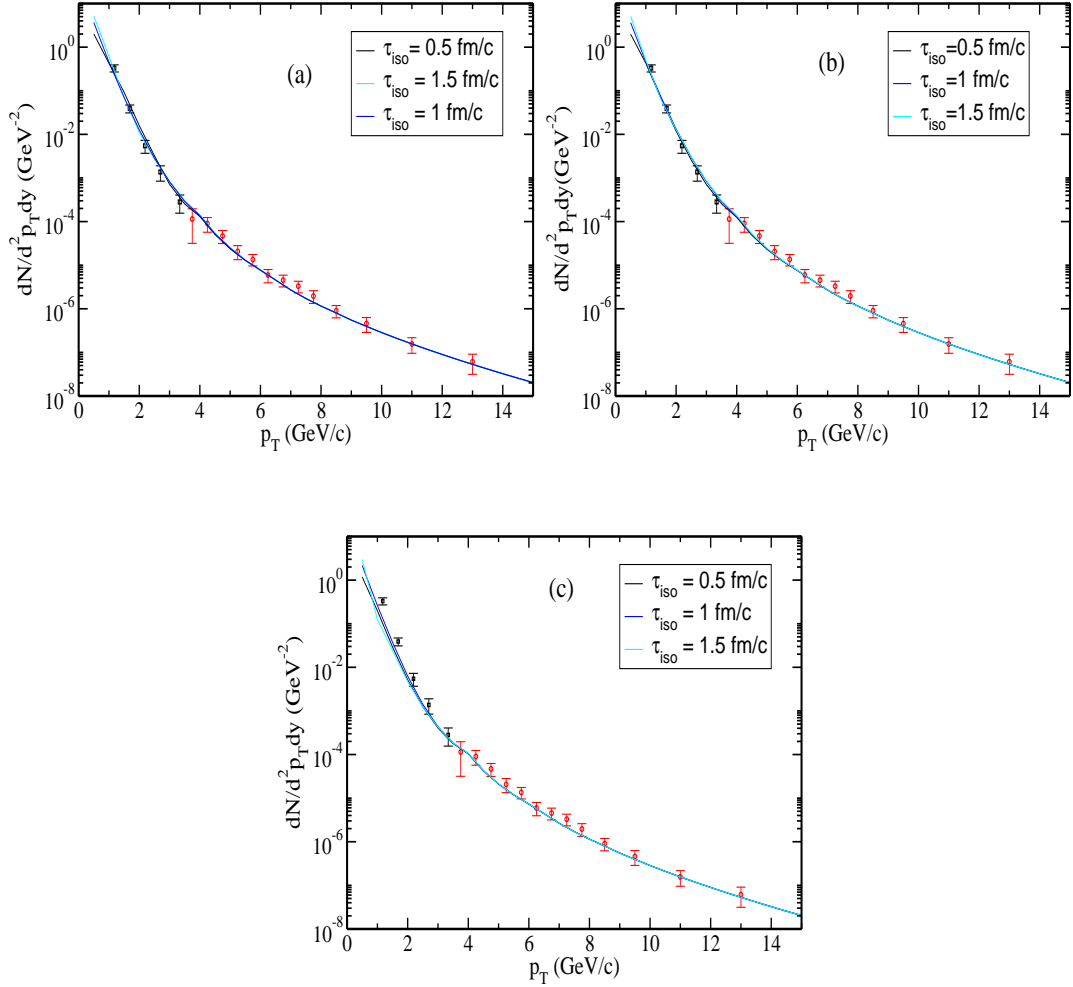

\begin{center}
\epsfig{file=fig4_440_alliso.eps,width=7cm,height=6cm,angle=0}
\epsfig{file=fig4_400_alliso.eps,width=7cm,height=6cm,angle=0}
\vskip 30pt
\epsfig{file=fig4_350_alliso.eps,width=7cm,height=6cm,angle=0}
\end{center}
\caption{(Color online) Photon transverse momentum distribution  with 
(a) Set I, (b) Set II and (c) Set III. The transition 
temperature is taken to be 192 MeV. Different lines show the yields for 
various values of $\tau_{\rm iso}$. The data points are taken 
from~\protect\cite{adler1,adler2}.}
\label{fig2}
\end{figure}
For any quantitative prediction of the expected total 
thermal photon yield the static photon rate, discussed in the 
previous section, has to be convoluted with the space-time 
evolution of the fireball. For the evolution scenario we propose 
the following. The system evolves anisotropically from 
$\tau_i$ to $\tau_{\rm iso}$ where one needs to know the time 
dependence of $p_{\rm hard}$ and $\xi$. We shall follow the work 
of Ref.~\cite{mauricio} to evaluate the $p_T$ distribution of 
photons from the first few Fermi of the plasma evolution. As the 
effect of transverse flow is pronounced in the late stages of the 
collisions we shall neglect this effect in the early stage. 
For $\tau > \tau_{\rm iso}$, the system is described by ideal relativistic 
hydrodynamics in (1+2)$d$~\cite{jpg50} with longitudinal boost 
invariance~\cite{jpg51} and cylindrical symmetry. As the
system becomes isotropic at 
$\tau = \tau_{\rm iso}$, $p_{\rm hard} (\tau_{\rm iso})$ 
and $\tau_{\rm iso}$ can be identified as the initial conditions, 
i. e., initial temperature and initial time for the hydrodynamic evolution. 
The time dependences of the anisotropy parameter $\xi$ and the hard scale
$p_{\rm hard}$ are taken from Ref.~\cite{mauricio}. The initial conditions 
for ideal hydrodynamics is obtained by the conditions,
\bea
T_i^{\rm hydro}&=&p_{\rm hard} (\tau_{\rm iso})
\nonumber\\
 \tau_i^{\rm hydro}&=& \tau_{\rm iso}
\eea
%
%


In our calculation, we assume a first-order phase transition beginning at the 
time $\tau_c (p_{\rm hard}(\tau_c)=T_c)$ and ending at $\tau_{H}=r_d\tau_c$ 
where $r_d=g_Q/g_H$ is the ratio of the degrees of freedom in the two 
(QGP phase and hadronic phase) phases. We shall consider two values
of the transition temperature $T_c = 192$ and 170 MeV~\cite{trans}.
The freeze-out temperature is fixed at $T_F =$ 120 MeV.
To cover the uncertainties in the initial conditions for RHIC energy, 
various combinations of $T_i$ ,$\tau_i$ consistent with 
the measured multiplicity ($dN/dy$) have been considered. We also vary 
$T_c$ to see its effect 
on the isotropization of the QGP. 

Therefore, the total thermal photon yield, arising from the present
scenario is given by,
\begin{equation}
\frac{dN}{d^2p_Tdy}=
\left[\int\,d^4x\, E\frac{dR}{d^3p}\right]_{\rm aniso} + 
\left[\int\,d^4x\, E\frac{dR}{d^3p}\right]_{\rm hydro}, 
\label{yield_total} 
\end{equation}
where the first term denotes the contribution from the anisotropic
QGP phase and the second term represents the contributions 
evaluated in ideal hydrodynamics scenario.

\subsection{Initial conditions and equation of state (EOS)}

To cover the uncertainties in the initial
conditions for a given beam energy, we consider three sets of
initial conditions, 
(I) $T_i=440$ MeV, $\tau_i=0.1$ fm/c
(II) $T_i=400$ MeV, $\tau_i=0.2$ fm/c, and
(III) $T_i=350$ MeV, $\tau_i=0.25$ fm/c
which are consistent with $dN/dy \sim 1100$ measured at RHIC 
energies. The initial energy density and  radial velocity 
profiles are taken as:
\be
\epsilon(\tau_i,r)=\frac{\epsilon_0}{1+e^{(r-R_A)/\delta}}
\label{enerin}
\ee
and
\be
v(\tau_i,r) = v_0\left[1-\frac{1}{1+e^{(r-R_A)/\delta}}\right]
\label{vrin}
\ee
We also need the EOS to solve the hydrodynamic equations.
Bag model type EOS has been used for QGP.
For EOS of the hadronic matter all the
resonances with mass $< 2.5$ GeV $/c^2$ have been 
considered~\cite{bedang}.

It is to be mentioned that in our case $\tau_{\rm iso}$ is always less than 
$\tau_c$ and we switch on
the transverse expansion at $\tau_{\rm iso}$ as the effect of transverse 
expansion in the very early stages is found to be negligible. Therefore,
for $\tau >\tau_{\rm iso}$ the energy density and the other thermodynamic
variables are functions of $r$ and $\tau$. The critical energy density
corresponding to the quark-hadron phase transition is, thus, also a contour
in the $(r,\tau)$ space. 

\section{Results}
 
We will now discuss contributions to the total photon yield
due to medium photon spectrum from anisotropic QGP with initial
conditions that might be achieved at 
RHIC. In what follows we shall consider the 
{\em free-streaming interpolating} 
model ($\delta =2 $). The results for 
{\em collisionally-broadened interpolating} 
model ($\delta =2/3$) are described in Ref.~\cite{lusakaprc}. 
In Fig.~(\ref{fig1}) we present the
photon yield due to Compton and annihilation processes in the mid 
rapidity ($\theta_{\gamma}=\pi/2$, $\theta_{\gamma}$ being the angle
between the photon momentum and the anisotropy direction) 
as a function of photon transverse momentum.
 Left (right) panel corresponds to $T_i= 0.440~(0.350)$ GeV and
$\tau_i= 0.1~(0.25)$ fm/c. In estimating these results, we have used
$\alpha_s=0.3$. Different lines in Fig.~\ref{fig1} correspond to 
different isotropization times, $\tau_{\rm iso}$. We clearly
observe enhancement of photon
yield when $\tau_{\rm iso}>\tau_i$. The enhancement of photon yield in 
the transverse directions ($y = 0$) is due to the fact that momentum-space 
anisotropy enhances the density of plasma partons moving at the mid 
rapidity~\cite{lusakaprc}.

Next we shall consider the total photon yield from various sources. 
This is displayed in 
Fig.~(\ref{fig2}) for different initial conditions as described in the 
text. It is seen
that the experimental data is well reproduced for all the three values of 
$\tau_{\rm iso}$ considered here. This is because of the following reason.
 For lower
values of $\tau_{\rm iso}$ the initial state momentum anisotropy leads to
lower yield as compared to higher values of $\tau_{\rm iso}$. But the 
initial temperature
($p_{\rm hard}(\tau_{\rm iso})$), required for hydrodynamic evolution
from $\tau_{\rm iso}$ onward is higher in the former case leading to 
higher yield.
These two competing effects are clearly revealed from the figure.
\begin{figure}
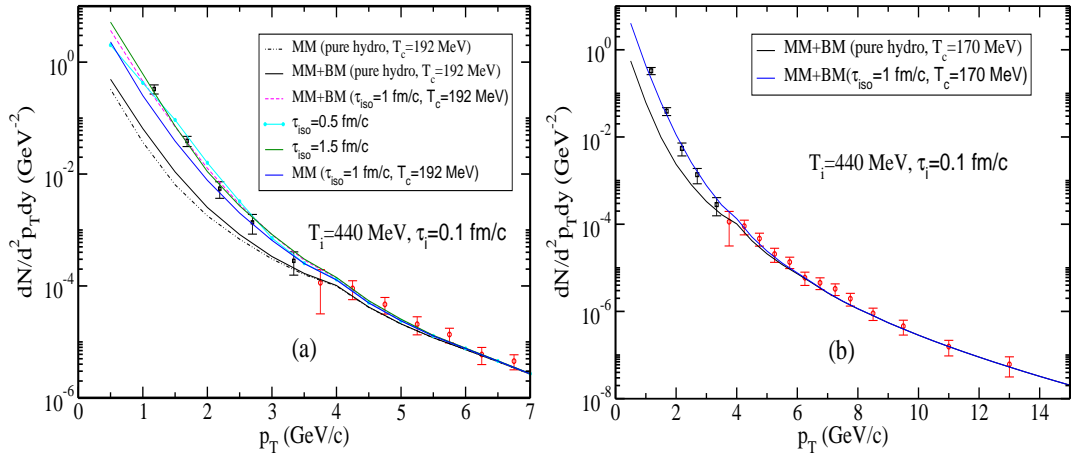

\begin{center}
\epsfig{file=data_440_192_inset.eps,width=7cm,height=6cm,angle=0}
\epsfig{file=fig2_440_Tc170.eps,width=7cm,height=6cm,angle=0}
\end{center}
\caption{(Color online) 
Photon transverse momentum distributions at RHIC energies with
initial condition Set I, for (a) $T_c$ = 192 MeV and (b) 170 MeV.}
\label{fig3}
\end{figure}
To show that the presence of initial state momentum anisotropy and the
importance of the contribution from baryon-meson reactions we plot the
the total photon yield assuming hydrodynamic evolution from the very 
begining as well
as with finite $\tau_{\rm iso}$ (right panel describes the total 
contribution with and without
the initial state momentum space anisotropy only for
$\tau_{\rm iso}$ = 1 fm/c) in Fig.~(\ref{fig3}). 
It is clearly seen that some amount of anisotropy is needed to reproduce 
the data. We note that the value
of $\tau_{\rm iso}$ needed to describe the data also lies in the range
1.5 fm/c$ \geq\tau_{\rm iso} \geq 0.5$ fm/c for both values of the transition
temperatures.    
. 
\begin{figure}
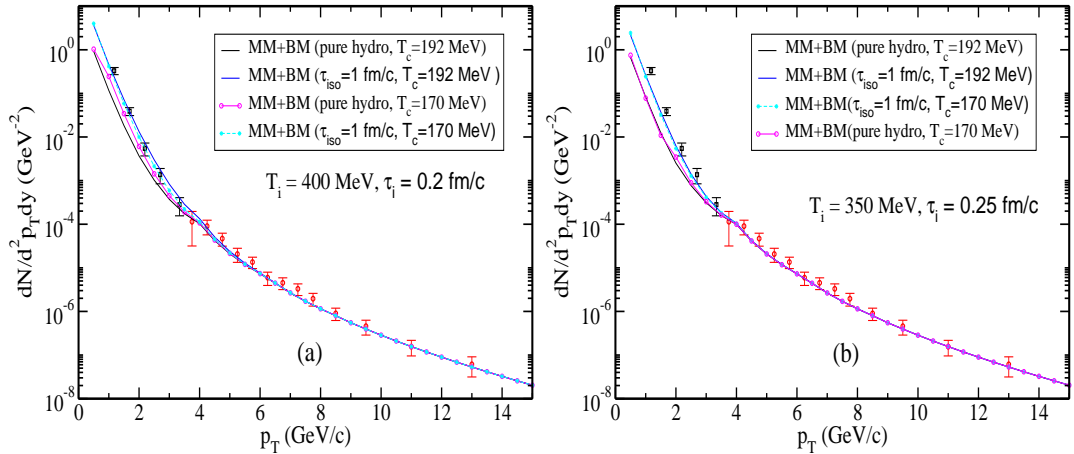

\begin{center}
\epsfig{file=fig3_400_allTc.eps,width=7cm,height=6cm,angle=0}
\epsfig{file=fig3_350_allTc.eps,width=7cm,height=6cm,angle=0}
\end{center}
\caption{(Color online) 
Total photon yield with (a) Set II and (b) Set III
with two values of transition temperatures, $T_c$ = 192 and 170 MeV.}
\label{fig4}
\end{figure}

To see the sensitivity of $\tau_{\rm iso}$ with the initial conditions 
we present the results with different initial conditions 
(Sets II and III) in Fig.~(\ref{fig4}) with $T_c$ = 192 (a) 
and 170 MeV (b). We again see that the  values of 
$\tau_{\rm iso}$ required to fit the data are in the range of 
$0.5 - 1.5$ fm/c and is almost independent of the transition 
temperature for a given initial conditions.

\begin{figure}
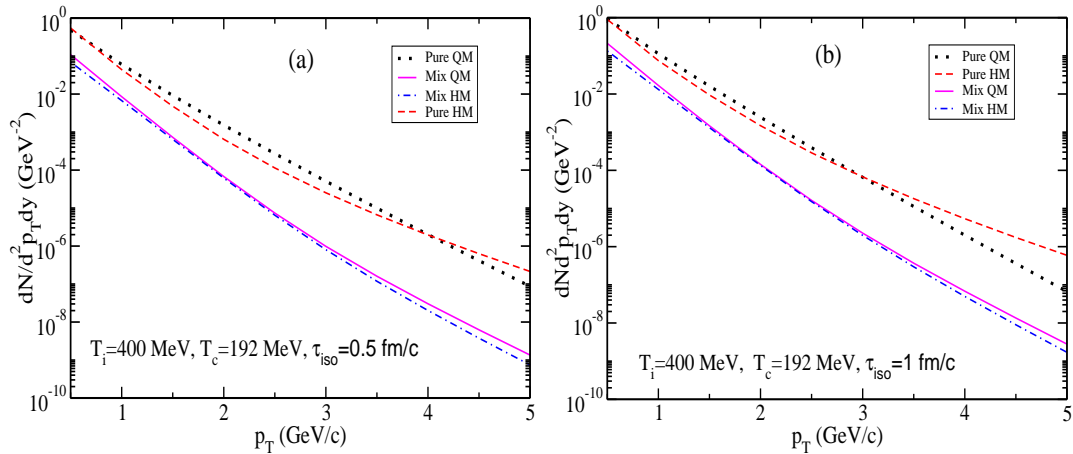

\begin{center}
\epsfig{file=fig2_report.eps,width=7cm,height=6cm,angle=0}
\epsfig{file=fig1_report.eps,width=7cm,height=6cm,angle=0}
\end{center}
\caption{(Color online) 
Contributions to total photon yield due to ideal hydrodynamics 
(second term of Eq.(8)) with Set II for (a)
$\tau_{\rm iso} = 0.5$ fm/c and (b) $\tau_{\rm iso} = 1$ fm/c 
for $T_c$ = 192 MeV. Here QM (HM) denotes quark matter (hadronic matter).}
\label{fig5}
\end{figure}

In order to see the hydrodynamic contributions to the total
photon yield we plot the second term of Eq.(8) in Fig.(\ref{fig5}). The 
left (right)
panel corresponds to $T_i^{\rm hydro} = 318 (348)$ MeV 
for $\tau_{\rm iso}$ = 1(0.5) fm/c, obtained by 
solving $T_i^{\rm hydro}=p_{\rm hard} (\tau_{\rm iso})$. We also 
note that $\tau_c = 2.28 (1.98)$ fm/c for 
$\tau_{\rm iso} = 1 (0.5)$ fm/c. It is found that because
of the transverse kick the low energy photons populate the intermediate
regime and consequently, the contribution from hadronic matter becomes
comparable with that from the hadronic matter destroying the window
where the contribution from QGP is supposed to dominate.

\section{Conclusions} 

To summarize, we have calculated total single photon transverse momentum
distributions by taking into account the effects of the pre-equilibrium
momentum space anisotropy of the QGP and late stage transverse expansion
on photons from hadronic matter with various initial conditions.  
To describe space-time evolution in the very early stage
we have used the phenomenological model described in Ref.~\cite{mauricio} 
for the time
dependence of the hard momentum scale ($p_{hard}$) and plasma 
anisotropy parameter ($\xi$). To calculate the hard photon contributions
we include the transverse momentum broadening in the initial hard scattering. 
The total photon yield is then compared with the PHENIX photon data. 
Within the ambit of the present model it is shown that the data can be 
described
quite well if $\tau_{\rm iso}$ is in the range of 0.5 - 1.5 fm/c for all 
the combinations of initial conditions and transition temperatures 
considered here. 
It is to be noted that the apparent hump observed in all the figures 
(except Fig.(5)) needs to be understood and we wish to discuss 
it in a subsequent paper. 
We conclude by noting that the isotropization time extracted from the 
PHENIX photon data is within the limit that is required to fit the 
other experimental observables, such as elliptic flow at RHIC using ideal 
hydrodynamics~\cite{Heinz}.

\end{document}